\documentclass[a4paper,10pt,twoside]{article}
\usepackage{amsmath} 
\usepackage{amssymb} 
\usepackage{amsthm}  
\usepackage{fancyhdr}
\usepackage{enumitem}
\usepackage[utf8]{inputenc}
\usepackage{setspace}
\usepackage{textcomp}
\usepackage[retainorgcmds]{IEEEtrantools}
\usepackage{url}
\usepackage[pdftex]{graphicx}
\usepackage[a4paper,twoside, hmarginratio={1:1}]{geometry}

\def\beqn{\begin{eqnarray}}
\def\eeqn{\end{eqnarray}}
\def\L{\mathcal{L}}
\def\MSSM{\text{MSSM}}
\def\tb{t_\beta}
\def\ma{M_{A^0}}
\def\Bsmu{B_s\to \mu^+\mu^-}
\def\Bsg{B\to X_s^*\gamma}
\def\chjm{\tilde{\chi}_j^-}
\def\chip{\tilde{\chi}_i^+}

\def\mh{m_h}

\def\Oh{\Omega h^2}
\def\mOmegas{\texttt{micrOmegas-2.4}}
\def\chio{\tilde{\chi}_1^0}

\def\sdt{s_{2\theta_t}}

\def\mneuto{m_{\tilde{\chi}_1^0}}

\def\Rgaga{R_{\gamma\gamma}}
\def\Rgagat{$R_{\gamma\gamma}\;$}

\begin{document}
\bibliographystyle{h-physrev}
\begin{titlepage}
\begin{center}

\vspace*{3cm}

{\Large {\bf Supersymmetric Higgses beyond the MSSM: An update with flavour and Dark Matter
constraints }}

\vspace{8mm}

{F. Boudjema and  G. Drieu La Rochelle}\\

\vspace{4mm}

{ LAPTh$^\dagger$, Univ. de Savoie, CNRS, B.P. 110,
Annecy-le-Vieux F-74941, France}

\vspace{10mm}

\today
\end{center}

\centerline{ {\bf Abstract} } \baselineskip=14pt \noindent

{\small 
Spurred by the discovery of a boson resonance at the LHC as the result of  the search for the
Standard Model Higgs, we pursue our investigation of the
properties and signatures 
of Higgses in an effective  supersymmetric scenario that goes beyond the usual MSSM. Such scenarios
were first introduced to alleviate the naturalness problem
of the MSSM Higgs and are found to have a very rich phenomenology that allows departures from the
Standard Model  in the production rate of the Higgs in many of the search channels.
We now include the constraints from flavour observables in particular the rare decays $\Bsg$ and
$\Bsmu$ including the recent measurement from LHCb. We also
address the issue of Dark Matter and its impact on Higgs physics. In particular, we incorporate the
latest data from XENON100 on the spin independent direct detection rates. These turn out to be
powerful constraints, especially if one also imposes that the observed thermal relic density is
obtained. We also study models 
with a low abundance that can more easily evade the direct detection rates. We study the impact of
the flavour and Dark Matter observables on the production rates of the Higgs at the LHC, and
their correlations in the diphoton, diphoton+jets and 4 leptons. We also comment on the other
channels.
}

\vspace*{\fill}

\vspace*{0.1cm} \rightline{LAPTh-039/12}

\vspace*{1cm}

$^\dagger${\small UMR 5108 du CNRS, associ\'ee  \`a l'Universit\'e
de Savoie.} \normalsize

\vspace*{2cm}

\end{titlepage}

\renewcommand{\topfraction}{0.85}
\renewcommand{\textfraction}{0.1}
\renewcommand{\floatpagefraction}{0.75}

\section{Introduction}
The July 4th 2012 announcement by both the ATLAS \cite{ATLASJuly4th, ATLASww2012} and
CMS \cite{CMSJuly4th} Collaborations of a $5\sigma$ resonance as a result for the search
of the standard model Higgs boson may well correspond to the discovery of the last missing piece of
the standard model, SM. The SM should then  be elevated to
the status of a theory  especially in view of the fact that the mass of this resonance is in accord
with the indirect limits from precision measurements. If
this particle turns out to be indeed the Standard Model Higgs with perhaps no new particles being
discovered, the naturalness argument that has motivated the
construction of so many beyond the SM (BSM) models will remain a mystery. These BSM constructions 
also aimed at providing a dark matter (DM) candidate for which the LHC might have provided some
circumstantial evidence. Probing the nature of the newly discovered resonance will 
certainly take time. Moreover the present data, though compatible with a SM Higgs interpretation
when all analyses are combined, seems to deviate from the prediction of the SM in some channels. In
particular the $2\gamma$ final
state channel points to a signal rate that is higher than what is predicted in the SM. Other
couplings, like the crucial coupling to $WW/ZZ$  require more data
taking. \\
As soon as the July results were made public, there has been a flurry of analyses aiming at fitting
the Higgs couplings in a model independent 
way \cite{Thfits_Higgs_July2012}. Other analyses concentrated on specific models. Most prominent
among the latter analyses was the status of the
MSSM \cite{Arbey:2012dq,Cao:2012yn} (MSSM for the Minimal Supersymmetric Standard Model).
Unfortunately the mass of the resonance, $\sim 125$ GeV, is very difficult to reconcile with
naturalness in the MSSM, see for
example \cite{cassel_bmssmhiggs_1103}. Moreover, unless one appeals to  quixotic
\cite{Carena:2012gp} choices of the parameters it is impossible for the $2\gamma$
rate to be higher in the MSSM than in the SM. Yet, it is hard to give up the idea of supersymmetry,
not only because of the DM candidate. One must then seek
models beyond the MSSM. The NMSSM (Next to MSSM) for example has been shown to fit better the
data \cite{ellwanger_1203,belanger_1203,Gunion:2012gc,Cao:2012yn}. Specific
extended versions with ultraviolet (UV) completion have also been proposed. It is therefore
important to follow an effective theory approach that encapsulates
the effects of a large class of specific UV completed scenarios beyond the MSSM and to 
parametrise the implications of some unknown model based on the symmetry of the low lying theory,
namely the MSSM. The main motivation for such an approach that keeps the same field content as
the MSSM, has been that the addition of a few operators in the Higgs
sector \cite{brignole,dine_bmssmhiggs_0707,antoniadis_bmssm_0708,
antoniadis_bmssmhiggs_0806,ponton_susy_higgs_0809,gdlr_higgs_1112,BMSSM125} alleviates very easily
the fine-tuning and naturalness problem. One no longer
requires heavy stops to have the lightest Higgs, $h$, weigh $125$ GeV. In fact, before the LHC Higgs
data of 2011 these generic BMSSM models accommodated a lightest
Higgs as heavy as $250$ GeV. The phenomenology of these models is very rich, since the properties of
the Higgses can change drastically. Such a set-up could serve
for an analysis of the two Higgs doublet model \cite{blum_1206,Azatov:2012wq,Craig:2012vn}.
Nonetheless the set-up is more restrictive, not only because of the
contribution of the higher dimensional operators to Higgs observables but also because  the
superfield implementation means  that the Higgsino sector is
affected. There are then implications on dark matter observables and, considering the link between
Higgs and heavy fermions, on flavour physics, in particular the
rare B-decays  $\Bsmu$ and $\Bsg$. LHCb \cite{lhcb_bsmu_1fb} has for example 
set new stringent constraints on $\Bsmu$.  As far as DM detection is concerned, July 2012 has also
seen XENON100 \cite{Xenon_july2012} set unprecedented bounds
on the rate of direct detection, while the last few years have witnessed a measurement of the relic
density of DM that has reached a precision of $3\%$. One has
therefore, no doubt, entered an exciting era in probing the details of symmetry breaking and
confront them with models that provide at the same time a dark matter candidate. \\
In view of the LHC results and the improvements that are expected in the coming months and years on
the reconstruction of the Higgs properties, an effective
theory approach that generalises  the usual MSSM and may encompass specific manifestations (extra
singlets \cite{basak_1204,Kyae:2012rv}, extra
triplets \cite{Delgado:2012sm}, U(1)', $\cdots$) is warranted. This paper is an update on our recent
detailed analyses \cite{gdlr_higgs_1112,BMSSM125} that took
into consideration all of the constraints on Higgs physics including the LHC 2011
data \cite{atlas_5fb_comb,cms_5fb_comb} the electroweak indirect precision
measurements and other constraints such as $t \to b H^+$. Like in our previous analyses we do not
aim at finding the best fits to the 2012 Higgs search/signal
data for the effective parameters of the higher order operators, not because they are numerous, but
because we consider such an exercise, considering the
experimental uncertainty, to be rather premature. Instead, in the 
signal region with an alleged Higgs of 125 GeV, we will compute the  possible signal strengths of
the different channels together with the correlations between
the different search channels. Although the mass for the alleged signal has narrowed around 125 GeV,
we still present our results for the range $122-128$ GeV. In 
this update we include the impact of the rare decays $\Bsmu$ and $\Bsg$ on the signal strengths. We
will then address the issue of dark matter, in particular the impact of the spin-independent direct
detection
constraints and then the relic density. Though some model dependence is introduced with DM, we will
see that the constraints  can shed
important light on the nature of DM. In a first stage we will assume that the BMSSM lightest
supersymmetric particle (LSP) accounts for all of DM and look how 
direct detection, in particular XENON100, restricts the parameter space and what consequences on the
Higgs signals it brings. In a second stage we investigate which of these scenarios do indeed provide
the observed relic abundance. In a third stage we review models where the abundance is low. Although
these models can not account for all of DM, the direct detection rates can be more easily evaded.
For such configurations we review the Higgs signal strengths. \\
Very little has so far been done as regards flavour and Dark Matter in the BMSSM and certainly not
from the point of view of the Higgs signal. Prior to the
Higgs signal results, 
flavour observables in the BMSSM have been studied in \cite{altmannshofer_bmssmcp_1107}. Dark
Matter, in particular the relic density, has been considered in
 \cite{cheung_bmssmdm_0903,gondolo_bmssmdm_0906,bernal_bmssmcosmo_0906,goudelis_bmssmdm_0912}. In
all these studies only the case of dim-5 operators has been
considered. In our study we include the full set of operators up to dim-6. \\
The paper is organised as 
follows : in section~2 a brief description of the model and the prominent experimental features are
presented. Some technical issues having to do with the calculation of the different observables are
reviewed in this section. Section~3 implements  the 
constraints from $\Bsmu$ and $\Bsg$ and study the consequences on the Higgs observables. With the
flavour constraints taken into account, we make the link with DM in section~4. First, we look at the
effect of direct detection as set by XENON100 (2012) assuming the model accounts for all of DM. We
then impose the bound set by the observed relic density. We finally consider models with an
abundance which is lower than what is observed. Section~5 summarises our conclusions.

 
\section{Description of the model}
Since the interested reader will certainly learn the details of the model in \cite{gdlr_higgs_1112}
and the references therein, we will only sketch a quick overview. The
model is within the effective theory approach where the effects of extra degrees of freedom beyond
those that describe the MSSM are taken into account through
higher order operators. The scale, $M$,  that enters these higher order operators is the heavy scale
of the New Physics (beyond the MSSM). We set this scale at 
\begin{equation}
M=1.5\text{ TeV},
\end{equation}
Since our low energy theory is supersymmetry (precisely the MSSM), $\L_\text{low energy}=\L_\MSSM$,
these effective operators will be products of superfields.
Given that we are not assuming anything on the UV  completion of the theory, those operators can be
any gauge and super Poincar\'e invariant
product of superfields. Since we concentrate on  Higgs phenomenology, we only consider operators
involving the Higgs superfields. In any case, in the same way that
radiative corrections to the Higgs in the MSSM have a very big impact, we suspect that these
operators can change the Higgs phenomenology in an important way.
As we \cite{gdlr_higgs_1112,BMSSM125}, and
others \cite{brignole,dine_bmssmhiggs_0707,antoniadis_bmssm_0708,
antoniadis_bmssmhiggs_0806,ponton_susy_higgs_0809}, have done so far, we include the dim-5
(superpotential) and dim-6 (K\"ahler) operators. 

They are the following :
\begin{IEEEeqnarray}{rCl}
W_{\text{eff}}&=&\zeta_1\frac{1}{M}\left(H_1.H_2\right)^2, \\
K_{\text{eff}}&=&a_1\frac{1}{M^2}\left(H_1^{\dag}e^{V_1}H_1\right)^2+a_2\frac{1}{M^2}\left(H_2^{\dag
}e^{V_2}H_2\right)^2+a_3\frac{1}{M^2}\left(H_1^{\dag}e^{V_1}
H_1\right)\left(H_2^{\dag}e^{V_2}H_2\right)\nonumber\\
&&+a_4\frac{1}{M^2}\bigl(H_1.H_2\bigr)\left(H_1^{\dag}.H_2^{\dag}\right)+\:\frac{1}{M^2}
\left(a_5H_1^{\dag}e^{V_1}H_1+a_6H_2^{\dag}e^{V_2}
H_2\right)\left(H_1.H_2+H_1^{\dag}.H_2^{\dag}\right).
\label{eq:eff_op}
\end{IEEEeqnarray}
$H_1,H_2$ are  Higgs superfields in the gauge basis with 
hypercharge  $Y_1=-1,Y_2=1$.  Supersymmetry breaking is introduced through the spurion
formalism \cite{antoniadis_bmssmhiggs_0910}. 
\begin{align}
\zeta_1&\longrightarrow\zeta_{10}+\zeta_{11}m_s\theta^2,\\
a_i&\longrightarrow
a_{i0}+a_{i1}m_s\theta^2+a_{i1}^*m_s\overline{\theta}^2+a_{i2}m_s^2\overline{\theta}^2\theta^2.
\end{align}
The approach, based on a supersymmetric set-up, assumes the physics beyond the MSSM to be
approximately supersymmetric and therefore $m_s$ is taken to be small
as compared to $M$. We take $m_s=300$ GeV.

The contribution of the dimensionless new parameters $a_{ij}, \zeta_{ij}$ to the Lagrangian
expressed in terms of the physical fields will be modulated by
powers of $m_s/M$ and 
 $\mu/M$, where $\mu$ is the usual supersymmetric Higgs mixing term. 
 At order $1/M^2$ the modulation enters as 
$(m_s/M)^2$, $(\mu/M)^2$ and $(\mu m_s/M^2)$ together with corrections of order $v^2/M^2$, $v$ being
the SM vacuum
expectation value. In order not to jeopardise the $1/M$ expansion we take $\mu=m_s$. In any case we
always impose a set of criteria in order to trust and
control the $1/M$ corrections, see \cite{BMSSM125}. In our analysis, the effective dimensionless
coefficients are varied within the range $[-1,1]$.

We would like to add a word of caution. Since the aim of this paper is to address the flavour (and
Dark Matter) issue one can not completely dismiss the
possibility of new operators that affect other sectors than the Higgs.We take the view here that
these effects are negligible compared to those emerging from
the Higgs sector. Our results will then be self-contained. 

\subsection{Parameter space}
Usually the low energy theory is fully specified. In our case $\L_\text{low energy}=\L_\MSSM$ and
therefore the parameters of the low energy model need to be
specified. 
In \cite{BMSSM125} we considered two set-ups, referred to as scenarios A and B. The reason behind
this choice lies in the importance of the stop sector and the
impact of the latter on the signature of the Higgs at the LHC. Therefore apart from the third
generation squarks, scenarios A and and B have the following
parameters. All soft scalar masses are set to $M_{{\rm soft}}=1$ TeV. As advertised earlier, the
Higgs $\mu$ parameter is set to 300 GeV. All
trilinear couplings are set to 0, except for the stop sector. 
The MSSM  parameters $\tb,\ma$ will be varied in the range
$$\tb\in[2,40],\quad\ma\in[50,450]\ \text{(GeV)}.$$ $\tb$ is the ratio between the expectation values in the Higgs doublets. 
$\ma$ is the mass of the pseudoscalar Higgs, $A^0$. The CP-even Higgses will be denoted as $h$ (the lightest) 
and $H$ (the heaviest). 
\\
The gaugino masses will only play a role when studying the DM. We set as benchmark 
$M_2$ (the $SU(2)$
gaugino mass) to 300 GeV, $M_1$ (the $U(1)$ bino mass) is
fixed by the universal gaugino mass relation
$M_1=\frac{5}{3}\tan^2 \theta_W M_2 \simeq M_2/2$, and $M_3=800$ GeV
(the $SU(3)$ gaugino mass), with $\cos^2\theta_W=M_W^2/M_Z^2$. When including DM observables we will
keep $M_2=\mu=300$ GeV fixed but will scan on $M_1$  in the range
$M_1 \in[50,300]$ GeV in order to generate all possible 
mixtures of higgsino-bino for the neutralino LSP. Indeed the nature of the LSP (bino, wino,
higgsino) is a key ingredient for dark matter observables.

The difference between scenario A and B is  in the third generation of squarks.
\begin{itemize}
 \item In model A we take $M_{{u3}_R}=M_{{d3}_R}=M_{Q_3}=400$ GeV, these are respectively the soft
masses for the up, down singlet and the doublet.  The
tri-linear stop mixing is $A_t=0$. This is a benchmark where the stops are light (as dictated by
naturalness considerations) but do not affect much the Higgs
loop couplings to $gg$ and $\gamma\gamma$ since they are mass degenerate.
 \item In model B we consider the case $m_{\tilde{t}_1}=200$ GeV together with $m_{\tilde{t}_2}=600$
 GeV and a maximal mixing in the stop sector with $\sin 2\theta_t=\sdt=1$. 
 This scenario exemplifies the role of stops in modifying the Higgs couplings to $gg$ and
$\gamma\gamma$ as compared to the SM
expectations. In \cite{BMSSM125} we covered a larger spectrum of $\tilde{t}_2$ and considered also
$\sdt=-1$. Fits to recent LHC data including flavour constraints in a MSSM set up with light stops have just appeared, 
see~\cite{Espinosa:2012in}. Note however, as we already pointed out in the introduction, such a {\em natural} set up 
has some tension with the {\em rather heavy} Higgs mass, which is not the case in the BMSSM. 
\end{itemize}

\subsection{Snapshot of the BMSSM}
Once the effective operators are plugged in the K\"ahler potential, the superpotential and the
susy-breaking potential we can derive the actual alterations to
the interactions of the physical fields themselves. The most  salient feature that has been
discussed thoroughly in the literature (see
 \cite{brignole,carena_bmssmhiggs_0909,antoniadis_bmssmhiggs_0910}) has to do with the substantial
increase in the lightest Higgs mass, $m_h$,  compared to the MSSM case.
In these scenarios $m_h$ can be raised up to 250 GeV. Although such high masses are no longer an
issue in view of the latest LHC results, this does show that contrary to the MSSM
a mass of 125 GeV 
for the lightest Higgs can be very easily attained, in particular without demanding too much from
the stops. What is important and is still of crucial
importance in view of the latest trends from the LHC is that the mixing and couplings of the BMSSM
Higgses can differ substantially from those of the Standard
Model. As we show in \cite{BMSSM125}, these deviations from the SM are however not haphazard despite
the relatively large number of parameters that the BMSSM
introduces. There exists strong correlations between different searches and signal channels. 

\subsection{Some technicalities on the computation}

The impact of these operators on flavour physics and Dark Matter has previously only been addressed
for the dim-5 operators arising from the superpotential, see
 \cite{altmannshofer_bmssmcp_1107} for the flavour observables and
 \cite{cheung_bmssmdm_0903,gondolo_bmssmdm_0906,bernal_bmssmcosmo_0906,goudelis_bmssmdm_0912}
for DM observable. However their impact on a Higgs with mass 125 GeV was not studied. The
implementation of both dim-5 and dim-6 operators on flavour physics
as well as the relic density and direct detection is performed here for the first time. As we
outline in a previous publication, the  implementation of the
effect of all the higher order operators is quite tricky. We perform this with an automated tool
from the superfield level to the physical states. We then pass
the newly created model file to external codes such as {\tt
micrOMEGAs} \cite{micromegas1,micromegas2,micromegas3,micromegas4} for the dark matter observables
for example. For our study on the Higgs observables all
these changes were a rescaling of the standard model  couplings.

\subsubsection{Flavour Observables}
For the flavour observables all what affects the Yukawa sector is of relevance and hence the
importance of these new contributions. 
For the calculation of observables which involve loop calculations one needs to be careful  that
these higher order contributions do not generate ultraviolet
divergences. One could imagine that all contributions from the higher order operators could be
naively counted as of a non renormalisable type. For example, in
the calculation 
of $\Bsmu$ enters the $\chjm\phi\chip$ vertex which contributes to the penguin diagram, $\chi^\pm$
are the chargino fields and $\phi=h,H,A^0$. At first sight the contribution from the higher order
operators exhibits
a new Lorentz structure containing derivatives on the chargino field. However,  it can be shown that
these new structures can be removed by using the equations
of motion. At the end,  the net effect  is fully taken into account by a rescaling of the MSSM
coupling. This then permits to easily adapt the calculation
performed in the MSSM, in this case \cite{bobeth_bsmumu,huang_bsmumu}.  In general all our
calculations of the flavour and Dark Matter observables take the codes
implemented in {\tt micrOMEGAs} as a skeleton. 

\subsubsection{Dark Matter}
Since our formulation stems from extra contributions involving Higgs superfields, the 
neutralino and chargino sector will be directly affected. This has an impact on the calculation of
Dark Matter observables in particular 
within a higgsino configuration.  The usual computation of the relic density and direct detection in
the BMSSM is particularly affected when higgsino and Higgs
are affected. For instance, for the relic density, processes involving 
Higgs final states such as $hA^0$ that occur when the neutralino is a mixed bino-higgsino get
corrected by as much as $30\%$ compared to the same MSSM point. In
the Higgsino case, the new operators allow for larger mass splitting between the neutralino and the
chargino. This helps evade more easily the LEP constraint on
the chargino. Another possibility that opens up is that co-annihilation with a stop are more
plausible than in the MSSM due to the fact that very heavy stops
are no longer necessary to obtain a Higgs with mass 125 GeV. For other novelties in the computation
of the relic density which however do not have a bearing on
our study of the Higgs see
\cite{goudelis_bmssmdm_0912,gondolo_bmssmdm_0906,bernal_bmssmcosmo_0906}. 

\subsection{Higgs observables at the LHC}
In order to use the results from the ATLAS and CMS collaborations,
we have used the following ratios
\begin{equation}
\label{RXX} R_{XX}=\frac{\sigma_{pp\rightarrow h\rightarrow
XX}}{\sigma_{pp\rightarrow h\rightarrow
XX}^{SM}}\qquad\text{and}\qquad
R_{XX}^\text{exclusion}=\frac{\sigma_{pp\rightarrow H\rightarrow
XX}}{\sigma_{pp\rightarrow H \rightarrow XX}^{\text{excluded
95\%}}},
\end{equation} where $XX$ denotes a particular final state
(say the inclusive $2\gamma$). $\sigma^\text{excluded 95\%}$
stands for the 95\% C.L. excluded cross-section reported by the
collaborations with the 2011 data (\cite{cms_5fb,atlas_5fb,cms_5fb_tautau}) : the reason why 2012
data for exclusion has not been used so far is that the most sensitive channels (notably
$\phi\to\bar\tau\tau, \;\phi=A^0,h, H$) have not been updated yet. In practice the $R_{XX}$ will be used in the
signal case, to compare with the best fit $\hat{\mu}$ -- of the so
called signal strength $\mu$-- given by the experiments. In
eq.~\ref{RXX}, $h$ in the BMSSM  will refer either to the lightest
or heaviest CP-even Higgs. $R_{XX}^\text{exclusion}$ will be used
in the no-signal case as a measure of the sensitivity of the
search, here $H$ stands for all Higgses not contributing to a signal
in the mass range $122-128$ GeV.  For $R_{XX}$ the most
important channels so far are the inclusive $2\gamma$, $ZZ \to 4l$  and the exclusive $2\gamma +2
jets$.\\

We simulate
the ratio $R_{\gamma\gamma+2j}$ as
\begin{equation} R_{\gamma\gamma+2j}=\frac{0.15\,\sigma_{\text{VBF}}+0.005\,\sigma_{gg\rightarrow
h}}{0.15\,\sigma_{\text{VBF}}^{SM}+0.005\,\sigma_{gg\rightarrow
h}^{SM}}\ \times\ \frac{BR_{\gamma\gamma}}{BR_{\gamma\gamma}^{SM}}
\end{equation}
We checked that this parametrisation of
$\sigma_{\gamma\gamma+\text{2 jets}}$ when folded in with the SM
cross sections for the LHC at $7$ TeV \cite{LHC_Higgs_cs2} and
taking into account the luminosity quoted by CMS reproduced quite
exactly the number of selected events given by CMS
 \cite{cms_5fb_gamgam}. We assume that this parametrisation that
was verified to be excellent for $m_h=120$ GeV still holds to a
very good degree in the range $122 < m_h < 128$ GeV. Note that with the 2012 data there exist
three different $2\gamma+\text{2 jets}$ channels (one for ATLAS and two for CMS) which correspond
to three different efficiencies. Given the high statistical uncertainty we will
however only consider one set of efficiencies as a representative of the channel.\\
We must note that we impose constraints from the electroweak precision data. Therefore the slightest
SU(2) custodial symmetry breaking effect is wiped out. The
models have therefore 
$R_{WW}/R_{ZZ}=1$.

\section{Impact of flavour on Higgs signals}
We first consider the impact of $\Bsmu$, $\Bsg$ and $(g-2)_\mu$ and look at how these observables
can constraint the rates for Higgs production in the different
channels

\paragraph{$\bullet \Bsmu$}~\\
We apply the latest bounds on $\Bsmu$ 
\begin{eqnarray*}
 \Bsmu&<&4.7\ 10^{-9}\qquad\text{LHCb \cite{lhcb_bsmu_1fb}}
\end{eqnarray*}

\paragraph{$\bullet \Bsg$}~\\ For  $\Bsg$ we take
\begin{eqnarray*}
 \Bsg&=&3.55\pm16\pm9\,\times10^{-4} \qquad\text{Heavy Flavour Averaging Group \cite{hfag_bsg}}
\end{eqnarray*}
where we have required the prediction to stay within two sigma deviations from the mean value. We
have taken the SM prediction to be $BR(\Bsg)=3.27\ 10^{-4}$
(see \cite{gambino_0805}). Note that any extra contribution beyond the SM 
is rather small.

\paragraph{$\bullet (g-2)_\mu.$} 
\begin{eqnarray*}
\Delta a_\mu=  a_\mu^{{\rm exp.}}-a_\mu^{{\rm th.}}=(2.8\pm0.8)\;10^{-9} \;\;\;
\cite{g-2_exp,g-2_th}.
\end{eqnarray*}

With the values of the MSSM parameters that we have taken (heavy sleptons) there is no effect from
$(g-2)_\mu$, either in terms of constraining the parameter space or 
alleviating the apparent $2\sigma$ discrepancy with the SM.   For $\Bsmu$, the effect is sensitive
to quite high values of $\tb$, $\tb \geq 20$ and small values of $\ma$, $\ma \leq 150$ GeV. This set
of parameter space is constrained by $\Bsmu$ is in fact  no longer
allowed by the Higgs exclusion limits set by the LHC itself in the analysis $\phi\to\bar\tau\tau$,
$\phi=h,H,A^0$ \cite{cms_5fb_tautau}, see later. Since the latter are
folded in our Higgs analysis, $\Bsmu$ does not add much. Note also that the effect of the dim-5
operators in particular are more important for small values of
$\tb$, therefore the BMSSM does not impact much more than the MSSM. 
$\Bsg$ is much more sensitive to the stop sector. We have observed that in our case most of the
supersymmetric corrections are brought by the Wilson operators
$O_7=(\bar s_L\sigma^{\mu\nu}b_R)F_{\mu\nu}$ : it is driven by the charged Higgs loop on one side
and the stops-charginos loop on the other. The former depends
on the value of $\ma$ and to a lesser extent on $\tb$. The latter shows a $\tb$-enhanced term whose
size is driven by $\sdt$ and $\Delta
m_{\tilde t}=m_{\tilde{t}_2}-m_{\tilde{t}_1}$. It is sensitive to the  sign of  $\mu$. In this study
we take $\mu > 0$.
Since the experimental value of $\Bsg$ is close to the SM prediction, it means that the
supersymmetric contribution must be quite small. This will drive us
either to a small $\tb$ region, a small mass splitting $\Delta m$ between stops or a small mixing
$\sdt \simeq 0$. The last two instances characterise model A. 

In model B, where we have a light stop $m_{\tilde{t}_1}=200$ GeV and a heavier one with 
$m_{\tilde{t}_2}=600$ (GeV), we expect $\Bsg$ to be more
constraining.


\subsection{Impact on Higgs observables}
The signal and correlations that we will show have $ 0< R_{\gamma \gamma} <4$. This is a very
generous band which  allows to read the predictions in a most transparent manner. The reader can
easily select a particular range. We prefer not to select a narrow range since  the
measurements on the different rates will evolve and get more precise. 

\subsubsection{Model A facing flavour}
\begin{figure}[!h]
\begin{center}
\mbox{
\includegraphics[scale=0.3,trim=0 0 0 0,clip=true]{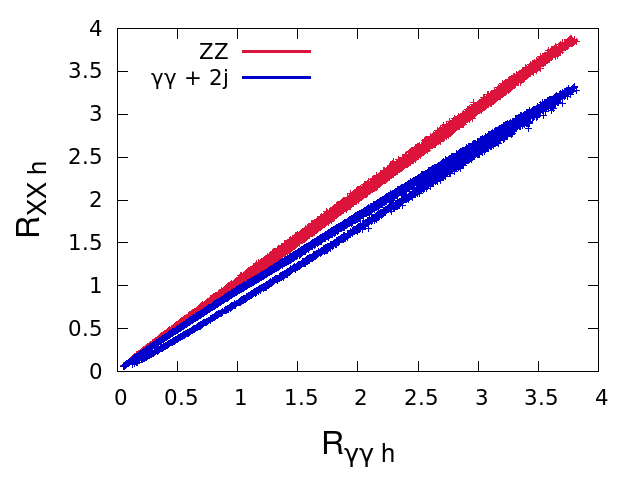}
\includegraphics[scale=0.3,trim=0 0 0 0,clip=true]{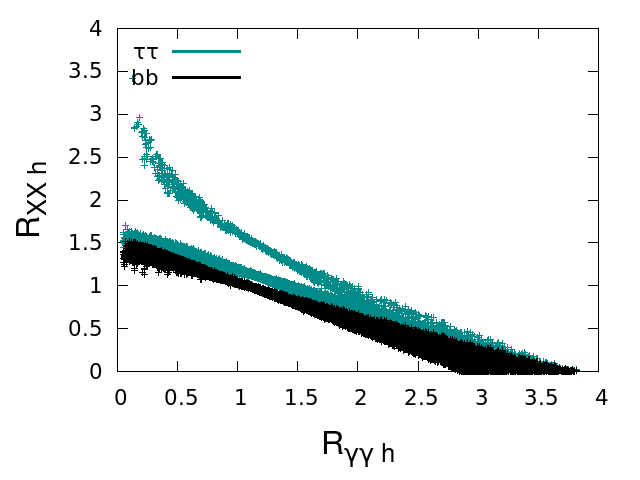}
}\\

\end{center}
\caption{\label{fig:flavour_allowed_A} {\em Allowed regions in  scenario A after applying the
flavour constraints for a signal with $\mh=125$ GeV. We plot here
the signal strengths and the correlations for 
a) Left panel:  $\gamma\gamma$ (x-axis), $ZZ$ (red points) and $\gamma\gamma$ + 2 jets (blue
points), b) right panel: $\tau \tau$ and 
$b \bar b$.}}
\end{figure}
In the case of model A with degenerate stops with mass of $400$ GeV, the correlations between the
signal strengths in the $\gamma \gamma$, $ZZ$ and $\gamma \gamma +2 j$ are
unaffected by the flavour constrained. Nor is the range of the signal strengths further reduced by
the flavour constraints. We note that in this particular case of degenerate stops the signals are
all strongly correlated with
$R_{\gamma \gamma} \sim R_{ZZ}$ and 
$R_{{\gamma \gamma} + 2j} < R_{\gamma \gamma}$. 
The correlations are shown in Fig.~\ref{fig:flavour_allowed_A}. 
For $R_{\gamma \gamma} \sim 2$ we can obtain the following ranges for the orther channels:
$R_{ZZ}=2-2.05$ and $R_{{\gamma\gamma}+2j}=1.6-1.8$, $R_{bb}=0.5-0.7$, $R_{\tau\tau}=0.6-1$. It is
important to point out that for this particular value, $R_{\gamma \gamma} \sim 2$, the $\tau \tau$
and $b \bar b$ need not be dramatically reduced. Much higher signal rates in the $\gamma \gamma$
channels are only possible with very much reduced of the latter two channels. 
\begin{figure}[!h]
\begin{center}
\mbox{
\includegraphics[scale=0.3,trim=0 0 0 0,clip=true]{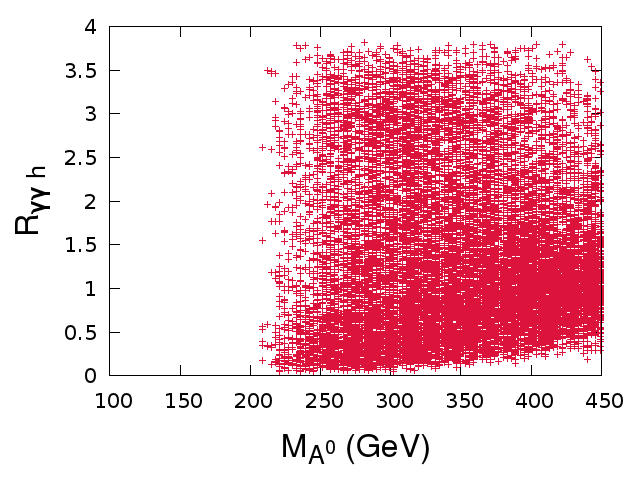}
\includegraphics[scale=0.3,trim=0 0 0 0,clip=true]{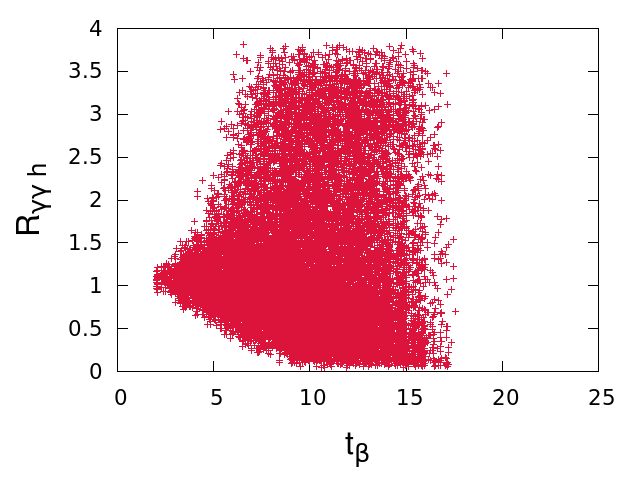}}
\end{center}
\caption{\label{fig:flavour_allowed_3} {\em Allowed region in $M_{A^0}$ versus $R_{\gamma \gamma}$
and  $\tb$ versus $R_{\gamma \gamma}$ in  scenario A after
applying flavour constraints. }}
\end{figure}

What is not visible in the projections of Fig.~\ref{fig:flavour_allowed_A} is the fact that the
flavour observables do eliminate quite a few configurations. Indeed we obtain the constraint
$\ma>200$ GeV and $\tb<20$, see
Fig.~\ref{fig:flavour_allowed_3}. These bounds come from the $\Bsg$ observable. Indeed at low $\ma$ the contribution of the charged
Higgs loop is enhanced. This can in principle be cancelled by
$\tb$-enhanced terms from the stop sector whose contribution has an opposite sign to the one of the
charged Higgs loop, however since there is practically no
effect of the stop sector in the model A the $\tb$ dependence of $\Bsg$ is very mild. Thus, in order
to reproduce the correct value for $\Bsg$ with $\ma<200$,
one would require $\tb>20$, a value which is already excluded by the $\phi\to\bar\tau\tau$ search at
the LHC. Since $A^0$ is pushed to $\ma>200$ GeV. This
means 
that the hypothesis of the heavy CP-even Higgs boson generating the signal is disfavoured. Such
possibility was entertained prior to applying the flavour constraints \cite{BMSSM125}.
Fig.~\ref{fig:flavour_allowed_3} reveals that for $\tb \sim 2$ we get 
$R_{\gamma \gamma} \sim 1$, while for $\tb \sim 5$ we obtain $0.5< R_{\gamma \gamma}< 2$. For higher
values of $\tb$ a much large range for $R_{\gamma \gamma}$ opens up. We
therefore see that, despite the many higher order operators, precision measurements on the Higgs
combined with flavour measurements can give a good measure of
$\tb$. We also see that in model A, values of $R_{\gamma \gamma} \sim 2$ are possible for any value
of $\ma > 250$ GeV, as long as $\tb > 5$. $R_{\gamma \gamma}<0.5$ would mean that the pseudo scalar
mass is lighter than $400$ GeV. 
\begin{figure}[!h]
\begin{center}
\includegraphics[scale=0.3,trim=0 0 0 0,clip=true]{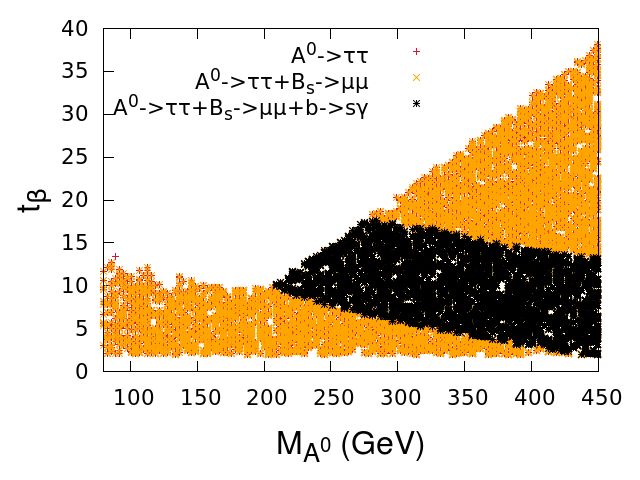}
\end{center}
\caption{\label{fig:modA_ma_tb} {\em Model A. Allowed points in the $\ma-\tb$ plane after imposing
the Higgs searches constraints (red), the $\Bsmu$ bound (orange) and the $\Bsg$ limit (black). }}
\end{figure}

Fig.~\ref{fig:modA_ma_tb} is very instructive. It reveals that $\Bsmu$ does not restrict the
parameter space once the LHC limit on the search $A^0 \to \tau \tau$ has been imposed. On the other
hand, $\Bsg$ carves out a significant region of parameter space.

\subsubsection{Model B facing flavour}
\begin{figure}[!h]
\begin{center}
\mbox{
\includegraphics[scale=0.3,trim=0 0 0 0,clip=true]{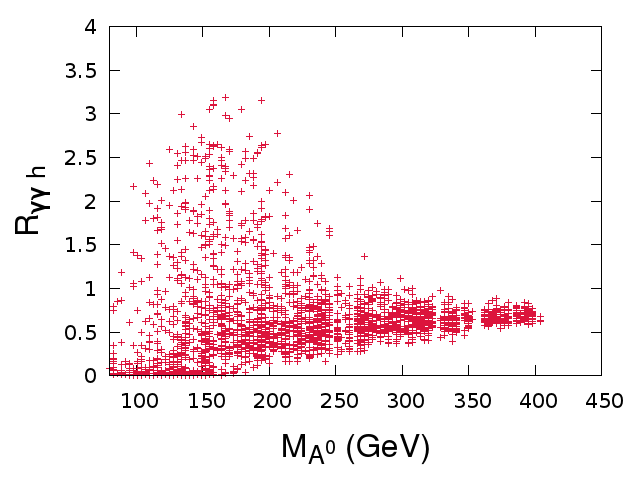}
\includegraphics[scale=0.3,trim=0 0 0 0,clip=true]{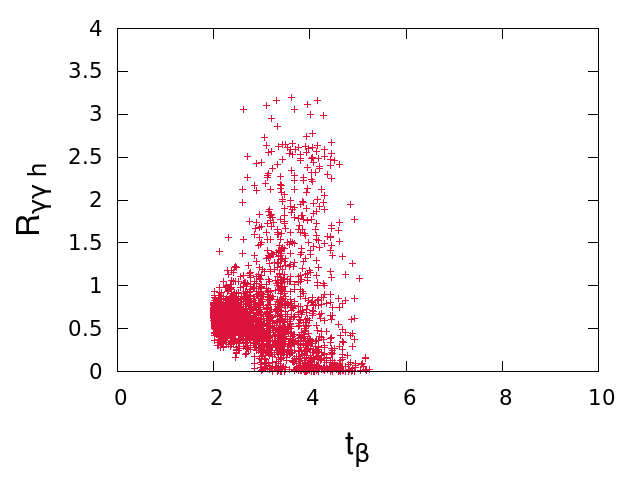}}
\caption{\label{fig:flavour_allowed_B_tb} {\em Allowed region in $M_{A^0}$ versus $R_{\gamma
\gamma}$ and  $\tb$ versus $R_{\gamma \gamma}$ in  scenario B after
applying the flavour constraints. }}
\end{center}
\end{figure}

\begin{figure}[!h]
\begin{center}
\mbox{
\includegraphics[scale=0.3,trim=0 0 0 0,clip=true]{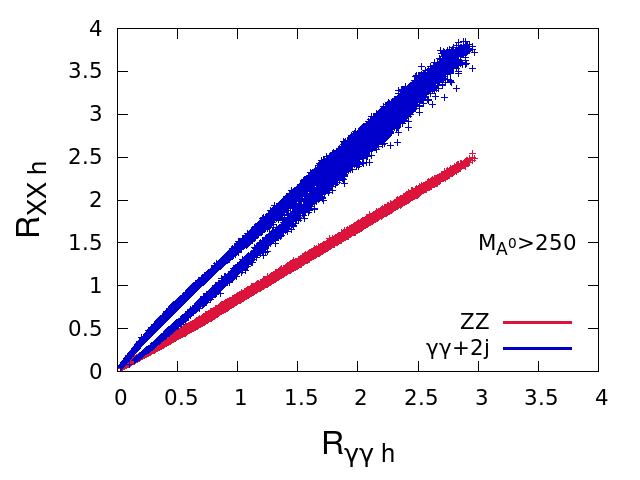}
\includegraphics[scale=0.3,trim=0 0 0 0,clip=true]{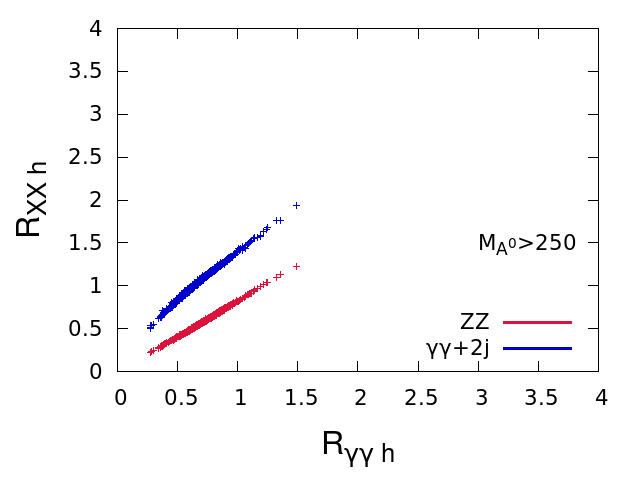}}
\mbox{
\includegraphics[scale=0.3,trim=0 0 0 0,clip=true]{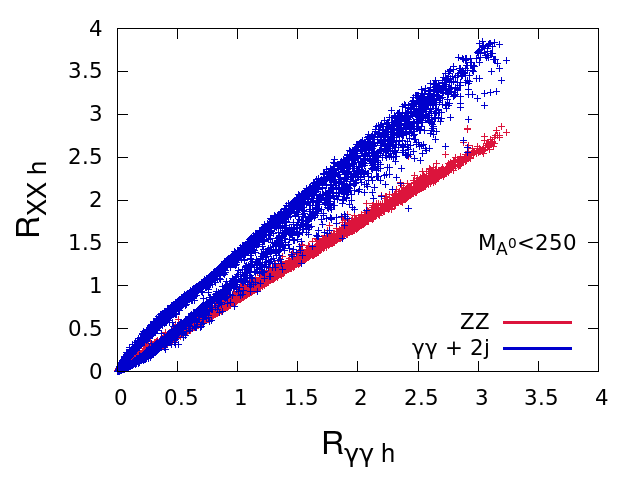}
\includegraphics[scale=0.3,trim=0 0 0 0,clip=true]{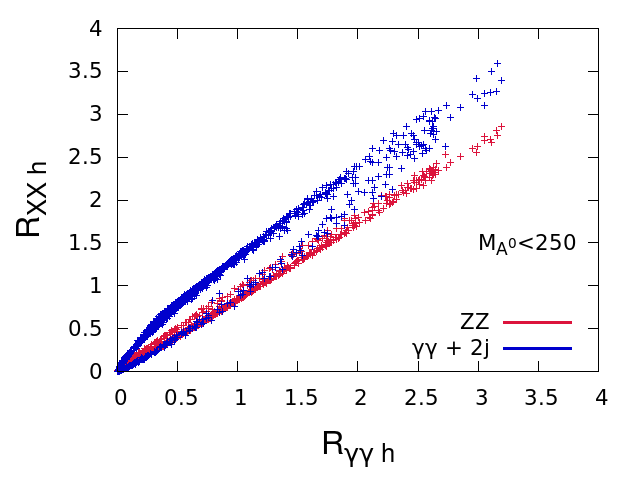}}
\end{center}
\caption{\label{fig:flavour_allowed_B} {\em Allowed regions in scenario B with $\sdt=1$ before (left
panels) and after (right panels)  applying the flavour
constraints. We distinguish the case of heavy ($M_A^0 > 250$ GeV) and light ($M_A^0 < 250$ GeV)
pseudoscalar masses. We plot here the features  of a
signal with $\mh=125$ GeV, that is to say the signal strength in the following channel :
$\gamma\gamma$ (x-axis), $ZZ$ (red points) and $\gamma\gamma$ + 2 jets (blue
points).}}
\end{figure}

\begin{figure}[!h]
\begin{center}
\includegraphics[scale=0.35,trim=0 0 0 0,clip=true]{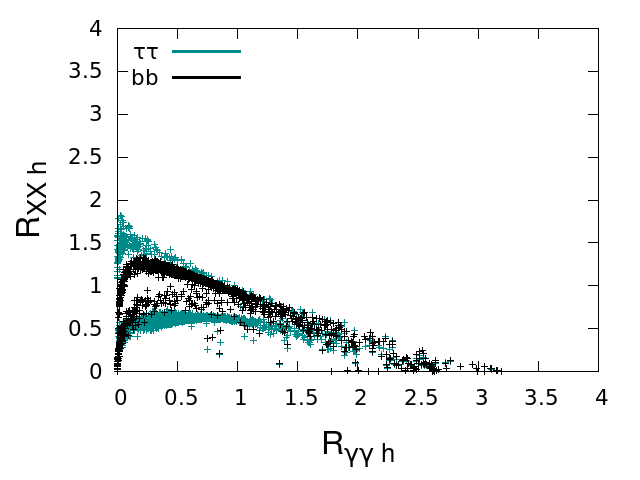}
\end{center}
\caption{\label{fig:modB_ma_tb}  {\em As in Fig.~\ref{fig:flavour_allowed_B} but for the  signal
strengths for $\tau \tau$ and $\bar bb$ channels  after flavour constraints are imposed.}}
\end{figure}

Model B has been introduced in order to obtain the hierarchy
$R_{\gamma\gamma+2j}>R_{\gamma\gamma}>R_{ZZ}$ by decreasing the contribution of the gluon fusion
with
respect to the $WW$ fusion. This is obtained through strong mixing in the stop sector and with one
of the stops relatively light. The underlying strong Yukawa
coupling of the stops can therefore 
have an important impact on $B$ observables in particular $\Bsg$. To illustrate this scenario we
take the case of maximal mixing with $\sdt=1$, more moderate
effects are obtained with smaller values of $\sdt$.  \\
The effect of the $\Bsg$ constraint is quite different from what we observed in model A. Indeed,
while the contribution of the charged Higgs is still the same,
with an important contribution for small  $\ma$, there is now a significant contribution from the
stop-chargino loop (since $\sdt=1$ and $\Delta m_{\tilde{t}}$
is non-zero) which is moreover $\tb$ enhanced. Since the latter has an opposite sign to the Standard
Model contribution, it will tend to decrease $BR(\Bsg)$ as
$\tb$ grows. The conclusion is twofold : first, the contribution of the charged Higgs can be
cancelled by the effect of the stop-chargino loop. Therefore, in
model B, on the one hand flavour constraints do not provide a lower bound for $\ma$, which is backed
up by the results shown in Fig.~\ref{fig:flavour_allowed_B_tb}. On the other hand, observe that
$\ma$ does not extend beyond $\ma> 400$ GeV, otherwise the
compensation between the charged Higgs contribution and the stop contribution in $\Bsg$ will not be
effective. Second, for the cancellation in $\Bsg$ to be effective and in
order to control the stop-chargino loop, $\tb$ is restricted
to be small ($\tb<5$) for any value of $\ma$. This is what is conveyed by
Fig.~\ref{fig:flavour_allowed_B_tb}. The range of $\tb$ is very much reduced compared to what we
found in Model A. Moreover for the largest allowed values of
$\ma$ $R_{\gamma \gamma}<1$. The fact that in this decoupling regime one does not recover the SM
value is  due to the reduction in $\sigma(
gg\to h)$ brought up by the stops.  \\
This restriction to small $\tb$ makes it difficult to obtain a  maximal suppression of the $g_{h\bar
bb}$ coupling that would lead to an  increase in $R_{\gamma \gamma}$, this is why in
Fig.~\ref{fig:flavour_allowed_B}, where we plot the allowed points in scenario B, we have much fewer
points with $R_{ \gamma\gamma}>2$ than before applying
the flavour constraints.  We note however that solutions with $R_{ \gamma\gamma}>2$ can not be
obtained with $\ma > 250$ GeV. With large values of $\ma$ decoupling will set in. Still,
there are configurations with  $R_{\gamma\gamma} \sim 2$ which exhibit the hierarchy 
$R_{\gamma\gamma+2j}>R_{\gamma\gamma}>R_{ZZ/WW}$. With $\ma < 250$ GeV we can attain
$R_{\gamma\gamma}=2$ together with
$R_{ZZ}=1.6-1.8$, $R_{\gamma\gamma+2j}=1.9-2.4$, $R_{bb}=0.3-0.6$ $R_{\tau\tau}=0.2-0.5$, see
Fig.~\ref{fig:modB_ma_tb}. Observe that $R_{\gamma\gamma} \sim 2$ corresponds to much lower rates
for the $b \bar b$ and $\tau \tau$ channels than in Model A, see Fig.~\ref{fig:modA_ma_tb}. With
$\ma > 250$ GeV, the increase in the {\em bosonic} final states ($\gamma \gamma, ZZ, \gamma \gamma
+2j$) is
reduced. While the correlations are maintained with $R_{\gamma\gamma+2j}/R_{\gamma\gamma}\sim 1.3,
R_{ZZ}/R_{\gamma\gamma}=0.8$, we now have $R_{\gamma\gamma} <1.5$ (for $\ma > 250$ GeV).

\section{Dark Matter}
Within the BMSSM, new features brought about by the extra operators have an impact also on non-Higgs
observables in particular the interaction of the higgsino
components. As such the properties of the lightest neutralino that could constitute a Dark Matter
candidate are affected, more so if the higgsino fraction is
important or else if the amount of mixing 
in the neutralino sector is important. Talking about the higgsino component, we should emphasize 
that in the BMSSM framework the value of  $\mu$ is not large. In this study we have fixed 
$\mu=300$ GeV. 
The results we have presented so far depend quite crucially on the value of $\mu$, in particular the
expansion in the effective operators is based on the ratio
$\mu/M$. The values of $M_{1,2}$ that determine the nature of the neutralino LSP have almost no
impact on the Higgs observables we have
studied. With $M_2 \sim 2 M_1$ and with $M_1 > 70$ GeV, invisible decays to dark matter neutralinos
is not possible and the contribution of charginos to $h \to
\gamma \gamma$ is negligible. Direct detection has a more direct connection with Higgs physics, due
to the contribution of Higgs exchange. The relic density requires the knowledge of
more parameters. For instance, lighter sleptons would have an important impact. This is the reason
why we first consider the constraint of direct detection on the
Higgs observables. In doing so we will, in a first stage,  assume that 
the density of Dark Matter is totally accounted for by neutralinos. Even if the relic density turns
out to be outside what is measured by WMAP, $\Omega h^2 \sim 0.1$,  one can always appeal to
non thermal scenarios which can bring the relic density to the desired experimental
value \cite{Gelmini:2006pw,Gelmini:2006mr}. We evoke this possibility only as
a way out not to include the impact of the relic density at this stage. In a second stage we compute
the relic density within a standard thermal cosmological
model and ask which scenarios can indeed be compatible with the correct relic density and pass the
direct detection constraint. Models that do not pass the cold
dark matter relic density constraint but give a value that is smaller than what is measured are
acceptable at the expense that the neutralino does not account
for the totality of DM in the Universe. Such possibilities are reviewed in a third stage. In this
case, given a spin-independent cross section,  the direct detection rate is smaller due to the
smaller
neutralino halo fraction. We assume that this fraction is the same as the one on cosmological scales
which is set by the relic density. The observed relic density $\
\Omega h^2$ is the result of the fit of $\Lambda{\rm CDM}$ whereas the direct detection rate is
proportional to the number density of the neutralinos
passing through the detector $\Omega_{\chi_1^0}$. In this case we reweight the result on the
spin-independent cross section and look at the effect on the Higgs
signal strength. \\
Let us at this point recall some important differences between model A and model B as regards the DM
candidate. In model B, the lightest stop weighs $200$ GeV, therefore the LSP must be lighter. As a
consequence, in these scenarios $M_1,M_2$ can not be taken very high, with $\mu=300$ GeV we can not
have $\mu \ll M_1,M_2$. 

\subsection{Direct Detection}
We have used the latest (July 2012) XENON 100 \cite{Xenon_july2012} results on the spin independent
cross-section of the dark matter candidate on the nucleus.
We have used  the routines of  \mOmegas\ \cite{micromegas2}.\\

In the framework we have chosen with squarks of the first and second generation being heavy and with
the possibility that the Higgses of the model, including the
pseudoscalar, can have mass below $300$ GeV, the rates for direct detection can provide a further
constraint on the parameters in the Higgs sector. The interplay
between direct detection and flavour in the context of the MSSM was emphasized in
 \cite{Belanger:2004ag}. 
The impact of direct detection depends also on the composition of the neutralino, of course. This
composition is determined by the values of the $\mu$ parameter
and the $U(1)$ ($M_1$) and $SU(2)$ ($M_2$) gaugino masses. The latter played practically no role in
the properties of the Higgs and the rates at the LHC. What
determines the cross section is the coupling of the higgses to the LSP neutralino and the coupling
of the Higgses to the quarks. These effects will naturally be
more important if the exchanged Higgs is not too heavy. For the coupling to quarks, high $\tb$ give
the largest effect.  We therefore expect
that adding the direct detection limit will constrain the Higgses couplings to the LSP. The latter
requires mixing between the Higgsino and the gaugino
($M_{1,2}$, in our case essentially $M_1$) components. \\

We first stick to the values of $M_1$ and $M_2$ that define models A and B. In this case with
$M_1\sim M_2/2=150$ GeV giving  $\mneuto \sim 146$ GeV for $\tb=20$, our benchmark
points barely make it. We find that very few points pass the new direct detection test for both
models. In fact as will be seen shortly, our benchmark choice for $M_{1,2}$ is borderline. This is
not difficult to explain. Indeed, although the
bino component is large ($90\%$), there is nonetheless about $10\%$ higgsino component. With the
latest limits from XENON100, such combined mixing and therefore such configurations
are almost ruled out both for model A and model B. The message is that XENON100 is now providing a
very strong constraint (also on many MSSM implementations). \\

We expect that in these scenarios reducing the amount of   higgsino-gaugino mixing will help. $M_1$
needs to be much further removed from $\mu$. We will therefore scan on $M_1$ so as to allow smaller
values than our benchmark $M_1=150$ GeV. In this study we do not entertain the possibility of
$M_{1,2} \gg \mu$.  In model B with $m_{\tilde{t}_1}=200$ GeV a DM candidate is not possible, while
for model A, $M_1$ would not extend above $400$ GeV and therefore we will be in a quite mixed
bino-higgsino configuration anyhow.   With $M_1<\mu$, our scan covers $M_1:70-300$ GeV (we keep
$\mu=M_2=300$ GeV). The lower value of $M_1$ is taken so as to avoid possible invisible decays of
the Higgs that we have not taken into account for our analysis (see however our
paper \cite{gdlr_higgs_1112}). \\

We first assume that neutralinos account for all of dark matter. The
flavour constraints are taken into account. \\

\begin{figure}[!h]
 \begin{center}
\mbox{
\includegraphics[scale=0.35,trim=0 0 0 0,clip=true]{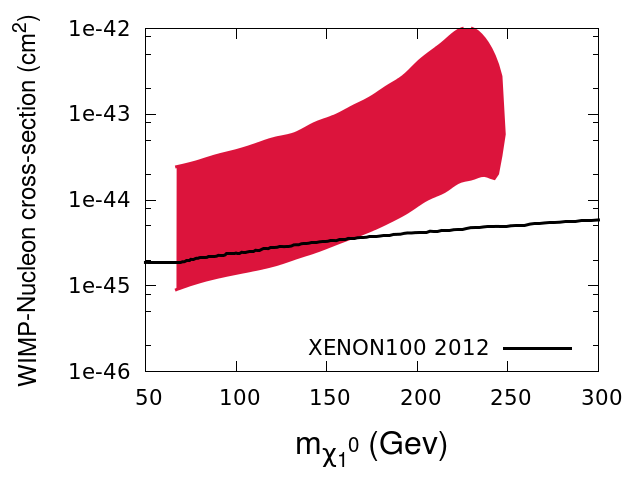}
\includegraphics[scale=0.35,trim=0 0 0 0,clip=true]{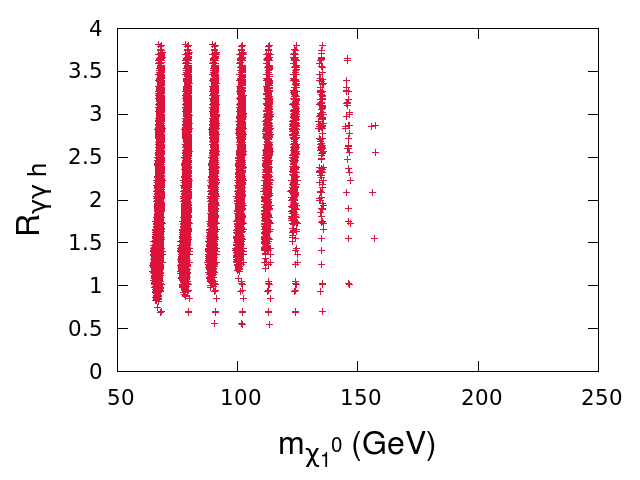}}
\mbox{
\includegraphics[scale=0.35,trim=0 0 0 0,clip=true]{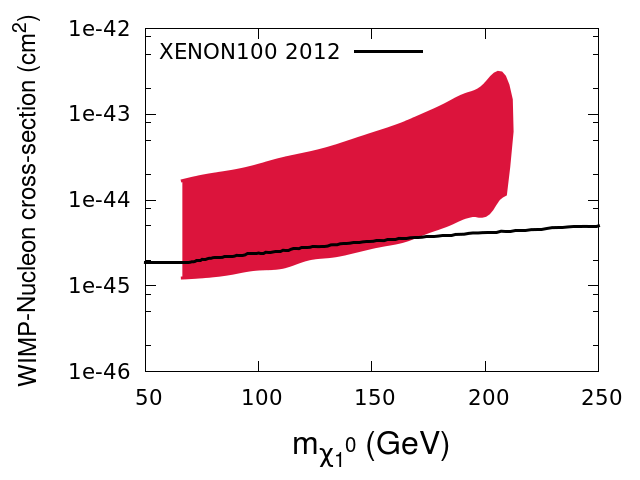}
\includegraphics[scale=0.35,trim=0 0 0 0,clip=true]{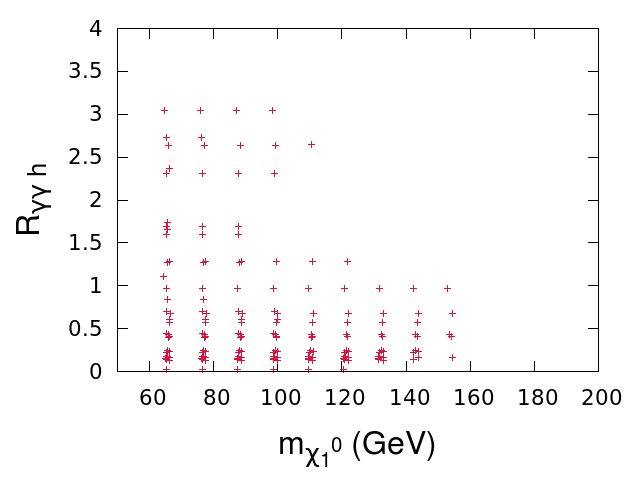}}
\end{center}
\caption{\label{fig:DD_allowedAB} {\em We show the allowed parameter space in
$m_{\chi_1^0}$-$\sigma_{\rm SI}$ after imposing XENON100 (2012) (left panels) and the impact on 
$R_{\gamma\gamma}$ (right panels). The upper (lower) plots are for model A (B).}}
\end{figure}

Looking at Fig.~\ref{fig:DD_allowedAB}   we see that we can find, in both model A and model B,
points that pass the XENON100 (2012) but only for neutralino
lighter than about $150$ GeV. Our benchmark point with $M_1=150$ GeV was indeed borderline. Many
configurations of the parameters including $\ma, \tb$, are therefore excluded. 
This shows that assuming that the models account for the bulk of DM, the new limit from
XENON100 (2012) are now extremely powerful. No doubt that future XENON1T \cite{Aprile:2002ef} which
will improve the sensitivity by at least an order of magnitude
will either soon discover such models or will exclude all of them. This is a conclusion that should
apply to all {\em natural} susy models with small enough
$\mu$ (see for example \cite{Baer:2012uy}).

The correlations between the different Higgs signal channels are, of course, unchanged. What is
important to check is whether the signal strengths are affected. We find that the only change
concerns Model A where direct detection now imposes $R_{\gamma \gamma}>0.5$,  see
Fig.~\ref{fig:DD_allowedAB}. After inspection we have found that  direct detection now cuts on the
small values of $\ma$ in particular those with largest $\tb$. In this case the couplings to $b\bar
b$ of the Higgs are not so large and hence the reduction in $R_{\gamma \gamma}$ is
more modest. In model A , $R_{\gamma \gamma}>1$ is possible (we even obtain $R_{\gamma \gamma}\sim 2
$) for all values of the neutralino mass in the considered range $60-150$ GeV. This is not the case
of Model B, where in the range $120 < \mneuto <150$ GeV, an enhancement of $R_{\gamma \gamma}$ is
not easy to find.

\subsection{Relic Density}
Combining the results of  the 7-year {\tt WMAP} data
 \cite{Jarosik:2010iu} on the 6-parameter $\Lambda$CDM model, the
baryon acoustic oscillations from {\tt SDSS} \cite{Percival:2009xn}
and the most recent determination of the Hubble
constant \cite{Riess:2009pu} one \cite{Komatsu:2010fb} arrives at
$\Omega h^2 = 0.1126 \pm 0.0036$, where $\Omega$
 is the density of cold dark matter normalised to the critical density, and $h$ is the Hubble
constant in units of $100$ km s$^{-1}$Mpc$^{-1}$. This experimental results is very precise with
only  3\% uncertainty. However, it has been shown, in particular
in supersymmetry, that such a precision was difficult to match on the theoretical side. Indeed the
loop corrections can easily be higher than 10\% on different
processes. Despite the efforts to account for those contributions (see references
 \cite{baro09,boudjema_chalons1,gdlr_dm_1108,Bjorn_review_rc}), it remains a
challenge and is so far not implemented in the code we have used, \mOmegas. We will thus impose the
value of the relic density within 15\% uncertainty :
\begin{equation}
\Oh=0.1126\pm0.016.
\end{equation}

Let us briefly sketch the main channels that enter the computation of  the relic density in our
scenarios:
\begin{itemize}
\item $\chio\chio\to f\bar f$ : this is the most frequent case when the lightest neutralino is
mainly bino-like. Though the cross section of this process is
usually too small to respect the relic density constraint, it can be enhanced by an $A^0$ resonance,
requiring $\ma \sim 2 \mneuto$.
\item $\chio\chio\to WW/ZZ$ : This occurs when the higgsino component is highest and the channels
are open. With $\mu=300$ GeV this takes place when 
$M_1 \geq \mu$. 
\item $\chio\chio\to WH/ZH/hA^0$ : this channel only opens up  for high masses, that is
$m_{\chio}>200$ GeV. 
\item $\chio {\tilde t}_1$ co-annihilation is also possible, in particular for Model B.
\end{itemize}

We now investigate whether in all scenarios we studied and that pass the direct detection constraint
one could still obtain the correct relic density within a
standard thermal cosmological model. We will start with model B where it is easier to illustrate why
we perform scans in steps of 10 GeV over $M_1$.

\subsubsection{Model B with the correct abundance}
\begin{figure}[!h]
\begin{center}
\includegraphics[scale=0.4,trim=0 0 0 0,clip=true]{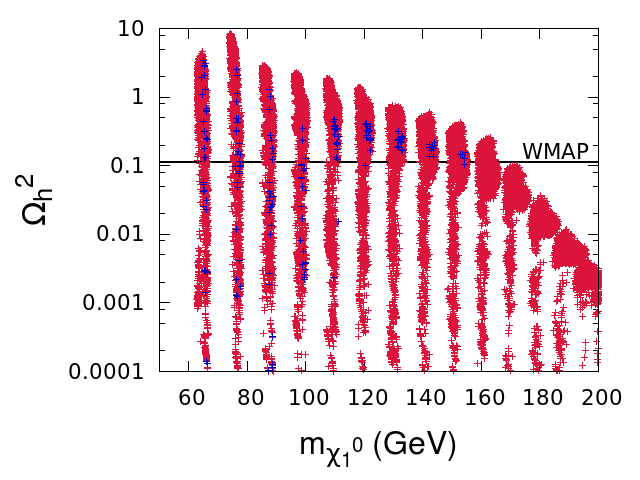}
\end{center}
\caption{\label{fig:relic_dd_B} {\em Model B. Values of the relic density (red/light grey) are
superimposed with those that have passed the XENON100 (2012)
limit(blue/dark grey). The WMAP bound is shown. The scan has been done by increasing the value of
$M_1$ in steps, rather than randomly (see text of why this was done)}}
\end{figure}
The good news is that it is possible to reproduce the correct thermal relic density and be in accord
with the latest measurement from XENON100 (2012) over the whole range  $60 < \mneuto <150$ GeV that
passed the XENON100 (2012) limit, Fig.~\ref{fig:relic_dd_B}.

\begin{figure}[!h]
\begin{center}
\mbox{
\includegraphics[scale=0.3,trim=0 0 0 0,clip=true]{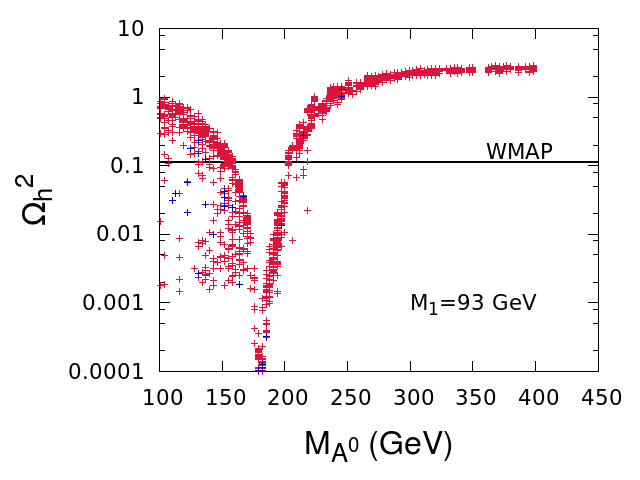}
\includegraphics[scale=0.3,trim=0 0 0 0,clip=true]{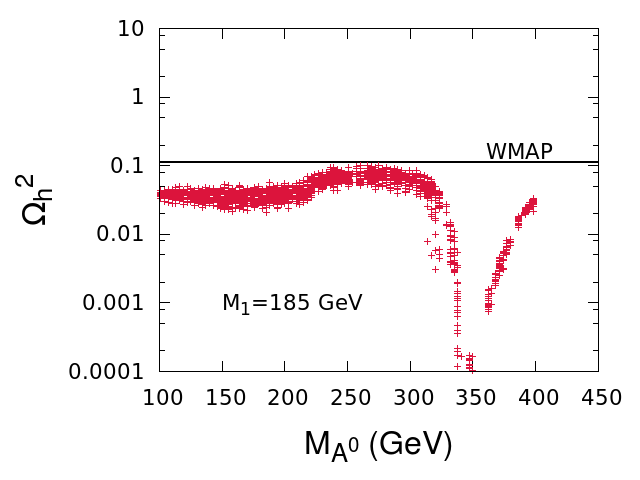}}
\end{center}
\caption{\label{fig:relic_res_B} {\em Model B. We show the result of a scan on $\ma$ on the relic
density as a function for $M_1=93$ GeV
(left panel) and $M_1=180$ GeV right panel. }}
\end{figure}

It is important to observe that  a scan over $M_1$ returns values for
the relic density that span a range over orders of magnitude in $\Omega h^2:10^{-4}-10$, with a
small subset that leads to the correct relic density and includes configurations in accord with
direct detection, see Fig.~\ref{fig:relic_dd_B}. Among the points that have passed the previous
direct detection limit some are associated either with 
an overabundance or an underabundance. The figure does not include configurations where the
spin-independent cross section is too large but which indeed corresponds
to an underabundance. We will deal with these scenarios next. Note  at this point that for $\mneuto
>160$ GeV, all scenarios represent underabundance. The figure illustrates the fact that as the mass
of the neutralino increases, and therefore the
higgsino component increases, annihilation of higgsino dominated LSP becomes more and more efficient
and the relic density drops.   The scan we have performed was done in steps of varying $M_1$ 
in order to reveal an important feature. As the value of $M_1$ increases and we enter the higgsino
domain, the strips in $M_1/m_{\chi^0_1}$ become wider for the
highest values (though still small) of the relic density. The pole-like structure (without a "head")
for low values of $M_1$ corresponds in fact to the
contribution of a Higgs resonance, $\ma \sim 2 \mneuto$. Around these tuned resonant regions  the
value of the relic density fluctuates vastly. An example of this is shown in
Fig.~\ref{fig:relic_res_B}. Obtaining the correct relic density could then be considered fine-tuned
with $\ma \sim 2 \mneuto$. As we reach the higgsino domain, these resonant processes become
irrelevant since other channels are more efficient. This  explains
the ``pole with the head" structure, see Fig.~\ref{fig:relic_dd_B}.   These regions correspond to
underabundance. 

The relic density constraint does not change the conclusions concerning the signal rates. We find
that in the range $ 60< \mneuto <120$ GeV we can have $\Rgaga>1$ ($\Rgaga\sim 2$ is possible here)
while for the rest of the allowed mass range  
$ 120< \mneuto <150$ GeV we have  $\Rgaga < 1$. Because the models which are retained are those with
$\ma \sim 2 \mneuto$ due to imposing the relic density constraint, these ranges can be converted to
ranges over $\ma$. Fig.~\ref{fig:flavour_allowed_B_tb} confirms the behavior of the 2-photon Higgs
signal strength. 

\subsubsection{Model B with an underabundance, a reappraisal of the direct detection}
\begin{figure}[!h]
\begin{center}
\includegraphics[scale=0.4,trim=0 0 0 0,clip=true]{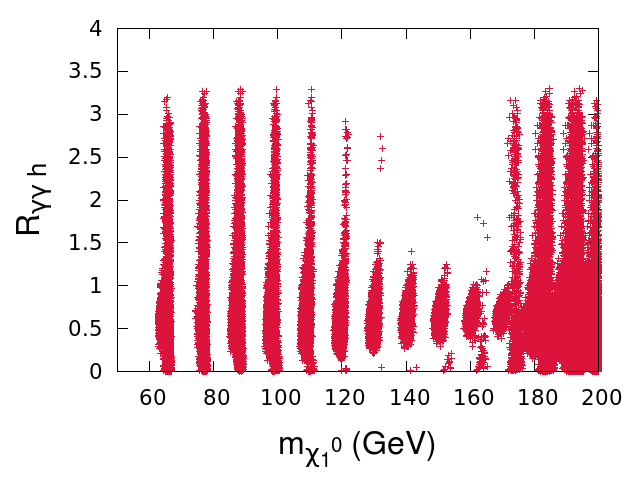}
\end{center}
\caption{\label{fig:underrelic_dd_B} {\em Model B : points with underabundance of the relic density
for which the modified XENON100 limit is respected.}}
\end{figure}

If neutralinos do not make up all of the dark matter, one can reconsider those scenario for which
the spin-indepent cross section seemed to high. Naturally configurations with $\mneuto > 160$ GeV
are now possible since the annihilation cross section for higgsinos are large and do not require the
contribution of a Higgs resonance contrary to scenarios with $\mneuto < 160$ GeV.
Fig.~\ref{fig:underrelic_dd_B} shows how $\Rgaga$ is affected. Up to $\mneuto \sim 160$ GeV we again
observe a gradual decline of the di-photon rate. In particular in the range of neutralino masses
between 120-160 GeV, this rate drops below that of the SM narrowing around a value of $0.5$. These
values can be interpreted in terms of the dependence of the \Rgagat as a function of $\ma$, see
Fig.~\ref{fig:flavour_allowed_B_tb} taking into account the fact that for these configurations $\ma
\sim 2 \mneuto$. On the other hand as soon as we enter the higgsino regime (and also co-annihilation
with stops), any value of $\ma$ will do to give a 
small enough relic density of neutralinos. In this case the di-photon rate is spread over a wide
range, in particular large enhancements are now possible. 

\subsubsection{Model A with the correct thermal abundance}
Many of the arguments that were detailed in the previous two sections for Model B can be invoked
when looking at the impact of the relic density on Model A. One common feature shared by  the two
models is that the dominance of the higgsino component in the calculation of the relic density kicks
in at about the same value of $\mneuto$, \textit{i.e.} $\mneuto \sim 160$ GeV. We have extended the
range of $\mneuto$ to about $250$ GeV, because the lightest stop is much heavier in Model A. At
around $\mneuto \sim 200$ GeV, we do not have the added contribution of the stop co-annihilation. 
\begin{figure}[!h]
\begin{center}
\includegraphics[scale=0.4,trim=0 0 0 0,clip=true]{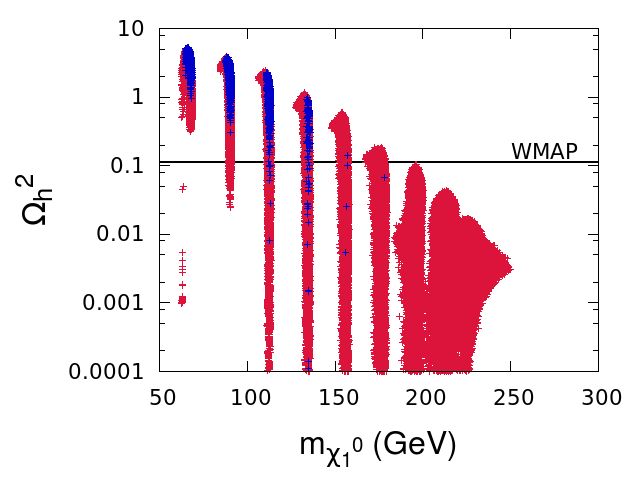}
\end{center}
\caption{\label{fig:relic_dd_A} {\em Model A. Values of the relic density (red/light grey) are
superimposed with those that have passed the XENON100 (2012)
limit(blue/dark grey). The WMAP bound is shown. }}
\end{figure}
As Fig.~\ref{fig:relic_dd_A} shows, the maximum value of the relic density drops steadily as the
neutralino mass increases. The pole like structures, 
indicative of an annihilation through a resonance, are still present. Recall however that contrary
to Model B, the flavour constraints have imposed $\ma > 200$ GeV, while allowing the larger range
$2-17$ for $\tb$. Yet we see a resonance like contribution that is much thinner around $\mneuto \sim
60$ GeV. This in fact is due to precisely the lightest Higgs. This said, although some of these 
configurations pass the  XENON100 (2012) constraint they do not simultaneously provide the standard
relic density, in fact these correspond to overabundances. Insisting on producing the present
abundance while abiding by the XENON limit, the masses of allowed neutralinos are in a narrower
range than in Model B : $100-160$ GeV.

\subsubsection{Model A with an underabundance, a reappraisal of the direct detection}
\begin{figure}[!h]
\begin{center}
\includegraphics[scale=0.4,trim=0 0 0 0,clip=true]{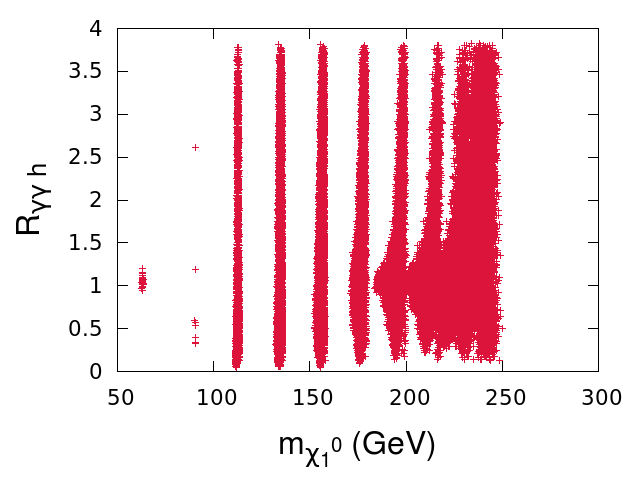}
\end{center}
\caption{\label{fig:underrelic_dd_A} {\em. Model A : points with underabundance  for which the
modified XENON100 limit is respected.}}
\end{figure}
We now allow that the neutralinos of Model A do not account for all of DM and assess which
neutralino masses are possible after imposing the flavour constraint and the direct detection limit,
see Fig.~\ref{fig:underrelic_dd_A}. The range $\sim 100$ GeV to $\sim 250$ GeV is now open, together
with the small island around $\mneuto \sim 60$ GeV that corresponds to annihilation through $h$. For
the latter, the di-photon rate is SM like. For the range $\mneuto=100-250$ GeV we span $0< \Rgaga
<3.5$. Contrary to model B, where the relic density proceeds through the pseudoscalar resonance for
$\mneuto:100-160$ GeV corresponding to $\ma > 200$ GeV, we reproduce   $0< \Rgaga <3.5$ as could
have been guessed from 
Fig.~\ref{fig:flavour_allowed_3}. For $\mneuto > 160$ GeV, the situation is similar to what we have
seen with Model B.

\section{Conclusions}
The results that the LHC Collaborations have announced in July 2012 are most probably pointing to
the discovery of a Higgs boson. If this is the SM Higgs boson, the naturalness argument upon which
one justified, for decades, the construction of new models for better explanation of symmetry
breaking would be most baffling. It is therefore important to seek 
whether the Higgs signals could be incorporated within a {\em natural} set-up, see for example the 
recent arguments in~\cite{ArkaniHamed:2012kq}. Supersymmetry would
be an ideal framework that could provide also a solution to the DM problem. However the relatively
heavy mass of the resonance discovered at the LHC suggests that the much studied MSSM will not be as
natural as wished, moreover if the excess in the di-photon signal is established, the MSSM will have
to be abandoned. Even before the LHC started taking data, there were signs of tension between the 
MSSM and naturalness. Keeping the supersymmetric framework but allowing for a more general set-up
had been advocated through the BMSSM to alleviate the problem. The series of analyses we have been
performing is to investigate whether the BMSSM is a viable alternative in the light of the new data.
Despite the large number of new parameters these effective models are rather constrained and
predictive. The study we have performed here aimed at reviewing what predictions and
correlations for the different signals of 
the Higgs are possible. While we eagerly await more precision on many of the Higgs rates, it is
important that one confronts these predictions with measurements concerning flavour and those that
address the DM observables. In this paper we considered two sets of scenarios. A BMSSM model with
degenerate stops at 400 GeV (Model A)  and a strongly mixed scenario in the stop sector with a
lightest stop at 200 GeV and a heavier one at 600 GeV, Model B. We find that the heavy flavour
observables, in particular $\Bsg$ (and to a lesser extent the new 
constraint from $\Bsmu$) delimit in an important way the parameter space of the general BMSSM. For
Model A, the pseudoscalar mass is restricted to $\ma > 200$ GeV while allowing for a relatively
large range for $\tb$, $\tb<20$. Such restrictions exclude the possibility that the signal at the
LHC could originate from the heavier CP-even Higgs. Despite these constraints, we still have
scenarios for a 125 GeV Higgs with rates higher in the di-photon signal than in the SM. Model A can
still give for example $R_{\gamma \gamma}\sim R_{ZZ}\sim 2$ while $R_{\gamma \gamma+2j}\sim 1.6$
while $R_{b \bar b} \sim 0.7$ and $R_{\tau \tau} \sim 0.7$. In Model B, the flavour constraints
could be considered even stronger. Indeed, for the maximal scenarios we have taken $\ma < 400$ GeV
and $\tb<5$. While the hierarchy $R_{\gamma \gamma+2j} > R_{\gamma \gamma} > R_{ZZ}\sim 2$ is
maintained it is possible to have $\Rgaga \sim 2$. The BMSSM models are also natural in the sense
that the
usual Higgs mixing parameter $\mu$ is not large. In 
our study this was set at $\mu=300$ GeV, justifying the approach of new operators associated with a
new physics at a scale $1.5$ TeV. This parameter is also important in defining the nature of the DM
candidate through the composition of the neutralino LSP. This composition depends on the weak
gaugino parameters $M_1,M_2$ which have, contrary to $\mu$ within the BMSSM, little impact on Higgs
observables. By looking at various $M_1$ versus $\mu$ hierarchies, we imposed the newly published
XENON100 limits.  These are very strong limits. We first assumed that the BMSSM LSP neutralino
accounts for all of DM. In both Model A and Model B, we find that only neutralinos with as little as
possible higgsino component pass the new limit. Therefore the LSP can not have mass above $160$ GeV.
The projected XENON1T will exclude all configurations with this assumption on the abundance. We then
ask whether the configurations that do indeed pass the XENON100 test have the correct relic density
as set by WMAP. We find that this is 
possible to achieve only if annihilation occurs through the pseudoscalar resonance with $\ma \sim
2\mneuto$. This could be considered as fine tuned. In turn, in model B,  for $120 < \mneuto <
150$ GeV even a small enhancement of the di-photon rate is difficult to achieve. For model A,
$\Rgaga > 1$ is possible over the whole allowed range $60 < \mneuto < 150$ GeV, in fact the
additional direct detection constraint imposes $\Rgaga > 0.5$. We have then asked how some
configurations can be rescued if we instead also accept scenarios with underabundance and
underdensity in the halo that lead to smaller direct detection rates even for large spin-independent
cross section. Masses of neutralinos up to the lightest stop mass are now possible. Apart from
neutralinos with mass in the range $120-160$ GeV in Model B where the Higgs signal strength is
small, in all other configurations we now span a large range of \Rgagat. The drawback in many of
these scenarios is that the BMSSM does not provide all of the observed DM, at 
least within a standard cosmological scenario. \\
Although we have not explored all the possibilities within the BMSSM implementations (\textit{e.g.}
we could have
considered larger values of $\mu$ together with a larger scale $M$, study more implementations of
the stop and stop mixings or the effect of a wino component), this study shows the importance of a
global study including Higgs, flavour and DM especially that new powerful data is pouring in. We
eagerly await the results from the projected XENON1T. Above all we keep a very close eye on the
upcoming analyses of the Higgs at the LHC. These include better measurements of as many channels in
the signal region and also further investigations of other mass regions that can probe the other
Higgses of these two doublet models. 
We stress once more that, despite the addition of many new operators beyond the usual MSSM, there
are strong correlations between the Higgs observables in the BMSSM. Direct searches for the stops
will also bring important information in these scenarios.





\bibliography{biblio}

\end{document}